\documentclass[twocolumn,showpacs,preprintnumbers,amsmath,amssymb,APSl,nofootinbib,superscriptaddress]{revtex4-1}

\usepackage{bm}
\usepackage{mathrsfs}
\usepackage{xcolor,color,graphicx,graphics}
\usepackage[all]{xy}
\usepackage{epsfig,subfigure}
\usepackage{latexsym,amssymb,amsmath,amsfonts} 
\usepackage[english]{babel} 
\usepackage[OT1]{fontenc}
\usepackage[latin1]{inputenc}
\usepackage{makeidx}
\usepackage{hyperref}
\usepackage{color,graphicx,graphics,wrapfig,epsf}
\usepackage{calligra}

\definecolor{red}{rgb}{1,0,0}

\def\+{^\dagger}

\def\<{\leftarrow}
\def\>{\rightarrow}

\def\({\left(}
\def\){\right)}

\def\a{\alpha}    
\def\m{\mu} \def\n{\nu}

\newcommand{\bi}{\begin{itemize}} 				\newcommand{\ei}{\end{itemize}}
\newcommand{\benu}{\begin{enumerate}} 		\newcommand{\enu}{\end{enumerate}}
\newcommand{\bd}{\begin{dinglist}{0}}     \newcommand{\ed}{\end{dinglist}}
\newcommand{\bfig}{\begin{figure}[htbp]}  \newcommand{\efig}{\end{figure}}
        			
\newcommand{\bc}{\begin{center}} 				  \newcommand{\ec}{\end{center}}
\newcommand{\be}{\begin{equation}} 				\newcommand{\ee}{\end{equation}}
\newcommand{\bsub}{\begin{subequations}}  \newcommand{\esub}{\end{subequations}}
\newcommand{\ben}{\begin{eqnarray}} 			\newcommand{\een}{\end{eqnarray}}
\newcommand{\ba}[1]{\begin{array}{#1}} 		\newcommand{\ea}{\end{array}}
\newcommand{\bea}{\begin{equation}\begin{array}{rcl}}
\newcommand{\eea}{\end{array}\end{equation}}

\begin{document}
\title{Mapping nonlinear gravity into General Relativity with nonlinear electrodynamics}

\author{Victor I. Afonso} \email{viafonso@df.ufcg.edu.br}
\affiliation{Unidade Acad\^{e}mica de F\'isica, Universidade Federal de Campina
Grande, Campina Grande-PB 58429-900, Brazil}
\author{Gonzalo J. Olmo} \email{gonzalo.olmo@uv.es}
\affiliation{Departamento de F\'{i}sica Te\'{o}rica and IFIC, Centro Mixto Universidad de Valencia - CSIC.
Universidad de Valencia, Burjassot-46100, Valencia, Spain}
\affiliation{Departamento de F\'isica, Universidade Federal da
Para\'\i ba, 58051-900 Jo\~ao Pessoa, Para\'\i ba, Brazil}
\author{Emanuele Orazi} \email{orazi.emanuele@gmail.com}
\affiliation{International Institute of Physics, Federal University of Rio Grande do Norte,
Campus Universit\'ario-Lagoa Nova, Natal-RN 59078-970, Brazil}
\affiliation{Escola de Ciencia e Tecnologia, Universidade Federal do Rio Grande do Norte, Caixa Postal 1524, Natal-RN 59078-970, Brazil}
\author{Diego Rubiera-Garcia} \email{drgarcia@fc.ul.pt}
\affiliation{Instituto de Astrof\'{\i}sica e Ci\^{e}ncias do Espa\c{c}o, Faculdade de
Ci\^encias da Universidade de Lisboa, Edif\'{\i}cio C8, Campo Grande,
P-1749-016 Lisbon, Portugal}

\date{\today}
\begin{abstract}
We show that families of nonlinear gravity theories formulated in a metric-affine approach and coupled to a nonlinear theory of electrodynamics can be mapped into General Relativity (GR) coupled to another nonlinear theory of electrodynamics. This allows to generate solutions of the former from those of the latter using purely algebraic transformations. This correspondence is explicitly illustrated with the Eddington-inspired Born-Infeld  theory of gravity, for which we consider a family of nonlinear electrodynamics and show that, under the map, preserve their algebraic structure. For the particular case of Maxwell electrodynamics coupled to Born-Infeld gravity we find, via this correspondence, a Born-Infeld-type  nonlinear electrodynamics on the GR side. Solving the spherically symmetric electrovacuum case for the latter, we show how the map provides directly the right solutions for the former. This procedure opens a new door to explore astrophysical and cosmological scenarios in nonlinear gravity theories by exploiting the full power of the analytical and numerical methods developed within the framework of GR.
\end{abstract}
\maketitle



\section{Introduction}

The exploration of new gravitational physics beyond General Relativity (GR) has always been plagued by technical difficulties. Even within GR, the highly nonlinear character of the equations of motion and their constrained structure  makes it difficult to address arbitrary dynamical situations both, from an analytical and a numerical perspective. Fortunately, important progress has been achieved on the numerical side which currently allows to confront observational data against model predictions  with extraordinary confidence \cite{Baumgarte:2002jm, Cardoso:2014uka, Font:2008fka, Bishop:2016lgv, Centrella:2010mx, Okawa:2014nda, Szilagyi:2015rwa}. A good example of this is represented by the recent observation of gravitational waves and their consistent interpretation in terms of binary mergers \cite{AbbotGW, AbbotPM, GBM:2017lvd, Abbott:2017dke, Abbott:2017gyy} (see \cite{Barack:2018yly} for a fresh review). Using that technical capacity to explore the predictions of theories beyond GR is not an easy task at all, as it would involve a substantial investment of time and human resources. The present work represents a further step into bridging the gap between a wide family of modified theories of gravity and the possibility of implementing well established analytical and numerical methods developed within the framework of GR for their analysis.

In a recent work \cite{Afonso:2018bpv}, some of us showed that for Ricci-based gravity theories (RBGs) in the metric-affine formulation
(no {\it a priori} relation imposed between the metric tensor and the affine connection), there exists a correspondence between the space of solutions of those theories and the space of solutions of GR. This has important technical implications, as one can define a problem in a given RBG theory, map it into GR, where it can be solved by standardized analytical or numerical methods, and then bring the obtained solution back to the original RBG theory via purely algebraic transformations, thus avoiding the need for developing specific methods for that particular RBG theory. In the present work we use that approach to explore spherically symmetric electrovacuum configurations and illustrate how the method works. As a test, we will recover some previously known solutions and will put forward the existence of certain symmetries between GR and the Eddington-inspired Born-Infeld  (EiBI)  gravity theory \cite{BIapp0,BIapp1}, which we use for concreteness and due to its recent interest in the literature regarding astrophysics and cosmology beyond GR (see \cite{BeltranJimenez:2017doy} for a recent review on this class of theories).

To proceed, we will first establish a correspondence between anisotropic fluid matter in GR and a generic RBG theory, particularizing it then to the EiBI model. This approach is followed because spherically symmetric electric fields can be naturally interpreted as anisotropic fluids. The previously obtained correspondence will thus allow us to address the electrovacuum problem in a simplified manner. This approach will put forward that, in general, the correspondence between GR and a nonlinear (in curvature) gravity theory induces specific nonlinearities in the matter sector of the GR frame. In other words, if we consider EiBI gravity coupled to Maxwell electrodynamics, the corresponding matter theory in the GR representation turns out to be a nonlinear theory of electrodynamics (NED). In general, any NED on the modified gravity side will be mapped into a different NED on the GR representation. For the particular case of EiBI gravity, we identify a family of NEDs which under the mapping to GR change as a M\"obius transformation. For this family it is also possible to determine the set of NEDs which remain invariant under the mapping. Understanding this aspect will shed useful light on some properties of the electrovacuum solutions of the EiBI gravity which were so far not fully understood.

The existence of this correspondence is particularly useful for the community working on astrophysical and cosmological applications of nonlinear models of matter, specially those in NEDs \cite{BI34,BIGR2,NEDregular1,NEDads3, Hassaine:2007py,Lemos:2011dq,EHGR2,NEDstructure8,NEDstructure4,Dymnikova:2015hka,NEDstructure3,NEDstructure5, NEDstructure6,NEDstructure7,NEDstructure10,NEDstructure11}. Indeed, in the latter case the GR solution is known in closed, exact form. This allows to find explicit solutions on the RBG side by solving algebraic equations rather than differential ones using the method here presented. The interest of this result is twofold. On the one hand, it breathes new life into many of such NED models, in particular, into those discarded on the GR side due to their lack of physical meaning, as the NED counterpart on the RBG/EiBI side may be of physical interest. On the other hand, by taking advantage of the full capacity of the analytical and numerical methods developed within GR, one can now explore in detail new astrophysical and cosmological applications of RBGs in less symmetric scenarios with electromagnetic fields, something previously hardly accessible due to the high nonlinearity of the RBG field equations.

The article is organized as follows: in section \ref{sec:II} we introduce the main elements of the mapping first presented in \cite{Afonso:2018bpv} for the RBG family of theories considered in this work, and outline the properties of their field equations in the Einstein frame. We proceed to describe how the mapping works for anisotropic fluids, and then particularize it to the EiBI gravity model. In section \ref{sec:III} we identify the particular class of anisotropic fluids electrovacuum fields correspond to, and work out the mapping for the (very general) family of M\"{o}bius-type NEDs. These results are particularized in section \ref{sec:IV} to EiBI gravity coupled to Maxwell electrodynamics. We find the corresponding solution identifying the associated NED on the GR side by direct application of the mapping. We conclude in section \ref{sec:V} with a summary and some perspectives for future research.

\section{Main elements of the mapping} \label{sec:II}

\subsection{Ricci-based gravity theories}

Consider the set of theories of gravity defined by
\begin{equation}\label{eq:actionRBG}
\mathcal{S}=\frac{1}{2\kappa^2}\int d^4x \sqrt{-g} \mathcal{L}_G\left[g_{\mu\nu},R_{\mu\nu}(\Gamma)\right]+\mathcal{S}_m[g_{\mu\nu},\psi_m]   \ ,
\end{equation}
where $\kappa^2$ is a constant with suitable dimensions, $g$ is the determinant of the spacetime metric $g_{\mu\nu}$, the scalar function $\mathcal{L}_G\left[g_{\mu\nu},R_{\mu\nu}(\Gamma)\right]$ is built out of traces of the object ${M^\m}_\n\equiv g^{\m\a} R_{\a\n}$, where the (symmetrized) Ricci tensor is defined as $R_{\mu\nu}(\Gamma) \equiv {R^\alpha}_{\mu\alpha\nu}(\Gamma)$, where $\Gamma \equiv \Gamma_{\mu\nu}^{\lambda}$ is the affine connection\footnote{In these theories, when the matter is represented by bosonic fields, the torsion (antisymmetric part of the connection) turns out to be a gauge degree of freedom and can be set to zero without physical consequences \cite{BeltranJimenez:2017doy,ACBO}.}. Here $\mathcal{S}_m=\int d^4x \sqrt{-g} \mathcal{L}_m(g_{\mu\nu},\psi_m)$ is the (minimally coupled) matter action, which is assumed to depend only on the spacetime metric and on the matter fields, collectively labelled by $\psi_m$. For the sake of this work we shall dub this family of theories as Ricci-Based Gravities (RBGs). RBGs are able to encompass a large variety of gravitational models such as (besides GR itself), $f(R)$, $f(R,R_{\mu\nu}R^{\mu\nu})$, or Born-Infeld inspired theories of gravity \cite{BeltranJimenez:2017doy}, all of which have attracted a great deal of interest in the recent literature \cite{FT10, CL11, Berti, NOO17}.

As we are working in the metric-affine (or Palatini) formalism, the field equations are obtained by independent variation of the action (\ref{eq:actionRBG}) with respect to metric and affine connection. The corresponding equations can be conveniently written under the form \cite{BeltranJimenez:2017doy,ACBO}
\begin{equation}\label{eq:GmnGeneral}
{G^\mu}_\nu(q)=\frac{\kappa^2}{|\hat{\Omega}|^{1/2}}\left[{T^\mu}_\nu-{\delta^\mu}_\nu\left(\mathcal{L}_G+\tfrac{T}{2}\right)\right] \ .
\end{equation}
where ${G^\mu}_\nu(q) \equiv  q^{\mu\alpha}R_{\alpha\nu}(q) - \frac{1}{2} {\delta^\mu}_{\nu} R(q)$ is the Einstein tensor of an auxiliary metric $q_{\mu\nu}$. This new metric fulfils the compatibility condition with the independent connection, {\it i.e.},
$\nabla_{\mu}^{\Gamma} \, q_{\alpha\beta}=0$ (thus $\Gamma_{\mu\nu}^{\lambda}$ is given by the Christoffel symbols of $q_{\mu\nu}$), while non-metricity is present for the spacetime metric, $Q_{\mu\alpha\beta} \!\equiv\! \nabla_{\mu}^{\Gamma} g_{\alpha\beta}\neq 0$. In the above expression the indices of the stress-energy tensor $T_{\mu\nu}=\frac{-2}{\sqrt{-g}}\frac{\delta (\sqrt{-g}\mathcal{L}_m)}{\delta g^{\mu\nu}}$ are raised with the space-time metric, ${T^\mu}_{\nu}\equiv g^{\mu\alpha}T_{\alpha\nu}$, so that $T \equiv g^{\mu\nu}T_{\mu\nu}$ denotes its trace. It is worth noting that the auxiliary metric $q_{\mu\nu}$ admits a nice interpretation in terms of an analogy between some condensed matter systems and gravitational physics  \cite{Lobo:2014nwa,Olmo:2015bha,Kroner,Kleinert} in which the nonmetricity tensor arises due to the existence of point-like defects in the underlying microscopic structure. In that analogy, $g_{\mu\nu}$ describes the physical (defected) geometry, while $q_{\mu\nu}$ represents an idealized structure without defects (vanishing nonmetricity).

Written in the form (\ref{eq:GmnGeneral}), the resulting field equations represent a set of second-order, Einstein-like equations for the  metric $q_{\mu\nu}$, where the right-hand side depends on the matter fields and, possibly, on $g_{\mu\nu}$. This is particularly useful from both a conceptual and an operational point of view, since in RBGs the relation between the auxiliary, $q_{\mu\nu}$, and spacetime, $g_{\mu\nu}$, metrics can always be written as
\begin{equation} \label{eq:qggen}
q_{\mu\nu}=g_{\mu\alpha}{\Omega^\alpha}_{\nu} \ .
\end{equation}
The \emph{deformation} matrix ${\Omega^\alpha}_{\nu}$ (hereafter vertical bars will denote its determinant) depends on the particular RBG chosen but, likewise $\mathcal{L}_G$, it can always be written  on-shell as a function of the stress-energy tensor, ${T^\mu}_\nu$ (and possibly on $g_{\mu\nu}$ as well). It should be pointed out that this $q_{\mu\nu}$-representation of the field equations highlights the fact that gravitational waves will propagate upon the background defined by $q_{\mu\nu}$, rather than the one defined by $g_{\mu\nu}$ (see \cite{BeltranJimenez:2017uwv} for an explanation in the case of Born-Infeld-type theories of gravity, and \cite{Jana:2017ost,Bazeia:2015zpa} for some related phenomenology).

An appealing feature of metric-affine theories as follows from the discussion above is that in vacuum, ${T^\mu}_{\nu}=0$, the deformation matrix becomes the identity. Accordingly, one has $q_{ \mu\nu}=g_{\mu\nu}$ (modulo a trivial re-scaling) and the field equations (\ref{eq:GmnGeneral}) of RBGs yield GR, possibly with a cosmological constant term depending on the form of $\mathcal{L}_G$. This implies that metric-affine RBGs, as defined by  Eq.(\ref{eq:actionRBG}) do not propagate extra degrees of freedom beyond the standard two polarizations of the gravitational field travelling at the speed of light (same as in GR). In turn, this allows these theories to successfully pass both solar system experiments \cite{Will:2014kxa,Olmo:2005hc,Olmo:2005zr}\footnote{For constraints of these theories from particle physics scattering experiments, improving the solar system constraints by a few orders of magnitude (depending on the particular RBG), see \cite{Latorre:2017uve}.}, and the recent gravitational wave observations from the LIGO-VIRGO collaboration related to equality of speed of propagation of gravitational and electromagnetic waves \cite{AbbotGW}, as well as the absence of additional polarization modes \cite{AbbotPM}, results which have either ruled out or placed strong constraints upon other proposals of gravitational schemes to extend GR \cite{Lombriser:2015sxa,Lombriser:2016yzn,Ezquiaga:2017ekz,Baker:2017hug, Sakstein:2017xjx,Creminelli:2017sry}.

We stress that while the left-hand side of Eq.(\ref{eq:GmnGeneral}) is a well defined function of the metric $q_{\mu\nu}$, the right-hand side depends nonlinearly on the matter fields and $g_{\mu\nu}$ via the functions $\mathcal{L}_G$ and $|\hat{\Omega}|$. Though the dependence on $g_{\mu\nu}$ can be explicitly solved in favour of $q_{\mu\nu}$ in specific cases, such as in homogeneous and isotropic cosmological models and in spherically symmetric scenarios, it is unclear that this can be done in more general and less symmetric configurations. In fact, from a numerical perspective, the inversion of that relation is likely to be computationally very expensive. Additional efforts are thus necessary to bring the above equations into a form that can be systematically worked out. Such is the purpose of this paper.

In order to show that the Einstein frame representation of the field equations (\ref{eq:GmnGeneral}) can be written without making any reference to the metric $g_{\mu\nu}$, one must consider that there exists a new matter source coupled to $q_{\mu\nu}$ such that the right-hand side of that equation can be written in the standard Einstein form
\begin{equation} \label{eq:GmnGR}
{G^\mu}_{\nu}(q)=\kappa^2 \bar{T}{^\mu}_{\nu} \ ,
\end{equation}
where $\bar{T}{^\mu}_{\nu}\equiv q^{\mu\alpha}\bar{T}_{\alpha\nu}$ represents the stress-energy tensor of a new set of matter fields of the same kind as the original ones, {\it i.e.}, fluids turn into fluids, scalar fields into scalar fields, and so on. There is a simple reason to establish this correspondence. Since the Einstein tensor on the left-hand side is conserved by virtue of the contracted Bianchi identity, $\nabla_\mu^q {G^\mu}_{\nu}(q)=0$, the right-hand side must also be conserved \emph{on-shell}. Since the right-hand side represents a certain kind of matter with deformations induced by the nonlinear gravity theory, in order to be conserved under the action of $\nabla_\mu^q $ it should be possible to express it in the form of the same kind of matter source (fluid, scalar, vector, \ldots), coupled to $q_{\mu\nu}$ in a way that guarantees its conservation. The simplest such a choice is, evidently, the form of a standard stress-energy tensor, which is the one we assume here. The equation above thus supports the convenience of the $q_{\mu\nu}$-representation of the field equations, which allows to transfer the problem of generating solutions in RBGs from solving differential field equations to algebraic ones (albeit non-linear, in general).

Following the above discussion, in order to establish a direct map between the spaces of solutions of RBGs and GR, by comparison between Eqs.(\ref{eq:GmnGeneral}) and (\ref{eq:GmnGR}), one must have
\begin{equation} \label{eq:mapping}
\bar{T}{^\mu}_{\nu}=\frac{1}{\vert \hat \Omega \vert^{1/2}}\left[{T^\mu}_{\nu}-{\delta^\mu}_{\nu}\left(\mathcal{L}_G+\frac{T}{2}\right)\right] \ ,
\end{equation}
which relates the effective stress-energy tensor generated by the matter coupled to the RBG, ${T^\mu}_{\nu}$, with the one coupled to GR, $\bar{T}{^\mu}_{\nu}$. Note that this map works irrespective of assumptions on symmetries of the problem or particular ansatze for the solutions. As we will see later, a canonical  matter Lagrangian (linear in the kinetic term) coupled to a given RBG generates, via the map, a noncanonical matter Lagrangian (nonlinear in the kinetic term) from the GR perspective. Thus, should one start with a canonical matter Lagrangian, the price to pay when going from a nonlinear gravity theory to its linear realization (GR) is to transfer the nonlinearities from the gravity sector to the matter sector. This will become apparent in the next sections.

\subsection{Anisotropic fluids}

For the sake of generality, let us consider RBGs in the action (\ref{eq:actionRBG}) coupled to an anisotropic fluid of the form
\begin{equation} \label{eq:Tmunufluid0}
{T^\mu}_\nu=(\rho + p_{\perp}) u^{\mu}u_{\nu}+p_{\perp} {\delta^\mu}_{\nu} + (p_r-p_{\perp}) \chi^{\mu}\chi_{\nu} \ ,
\end{equation}
where normalized timelike $g_{\mu\nu}u^{\mu}u^{\nu}=-1$ and spacelike $g_{\mu\nu}\chi^{\mu}\chi^{\nu}=+1$ vectors have been introduced, while $\rho$ is the fluid energy density, $p_r$ its pressure in the direction of $\chi^{\mu}$, and $p_{\perp}(r)$ its tangential pressure, in the direction orthogonal to $\chi^{\mu}$. Note that, in comoving coordinates, this fluid can be conveniently written as
\begin{equation} \label{eq:Tmunufluid2}
{T^\mu}_\nu=\text{diag}(-\rho,p_{r},p_{\perp},p_{\perp}) \ .
\end{equation}
(obviously standard perfect fluids are just a particular case of this stress-energy tensor with $p_{\perp}=p_r$). To work out the mapping in this case, consistently with the algebraic structure of this anisotropic fluid, we propose an ansatz for the matrix ${\Omega^\mu}_{\nu}$ of Eq.(\ref{eq:qggen}) under the general form
\begin{equation}  \label{eq:Omegafluid}
{\Omega^\mu}_{\nu}=\alpha {\delta^\mu}_{\nu} + \beta u^{\mu}u_{\nu} + \gamma \chi^{\mu}\chi_{\nu} \ ,
\end{equation}
where the explicit expressions of the functions $\{\alpha,\beta,\gamma\}$ (which, in general, will depend on $\{\rho,p_r,p_{\perp}\}$) can only be specified once a particular RBG model is chosen. This form for the matrix ${\Omega^\mu}_{\nu}$ is natural given that it is associated to a nonlinear function of ${T^\mu}_{\nu}$. Indeed, a power series expansion of ${\Omega^\mu}_{\nu}$ in terms of ${T^\mu}_{\nu}$ leads to the structure (\ref{eq:Omegafluid}) due to the orthogonality of the vectors $u^\mu$ and $\chi^\nu$, which prevents the existence of crossed terms.

Introducing these expressions into the RBG field equations (\ref{eq:GmnGeneral}) we get
\begin{eqnarray}
{G^\mu}_{\nu}(q)&=&\frac{\kappa^2}{\vert \hat{\Omega}\vert^{1/2}} \Big[\Big(\frac{\rho-p_r}{2}-\mathcal{L}_G\Big){\delta^\mu}_{\nu}  \nonumber \\
&+&(\rho+p_{\perp})u^{\mu}u_{\nu} +(p_r-p_{\perp}) \chi^{\mu}\chi_{\nu} \Big] \ .
\end{eqnarray}
Assuming now the existence of another anisotropic fluid on the GR side, defined also by Eq.(\ref{eq:Tmunufluid0}), but with new functions $\{\rho^q,p_r^q,p_{\perp}^q\}$, {\it i.e.},
\begin{equation} \label{eq:Tmunufluid1}
\bar{T}{^\mu}_{\nu}=(\rho^q + p_{\perp}^q) v^{\mu}v_{\nu}+p_{\perp}^q {\delta^\mu}_{\nu} + (p_r^q-p_{\perp}^q) \xi^{\mu}\xi_{\nu} \ ,
\end{equation}
for new timelike, $q_{\mu\nu}v^{\mu}v^{\nu}=-1$, and spacelike, $q_{\mu\nu}\xi^{\mu}\xi^{\nu}=+1$ vectors, then the mapping equations (\ref{eq:mapping}) become
\begin{eqnarray}
p_{\perp}^q&=&\frac{1}{\vert \hat{\Omega} \vert^{1/2}} \left[\frac{\rho-p_r}{2} -\mathcal{L}_G \right] \label{eq:mapflu1} \\
\rho^q+p_{\perp}^q&=&\frac{\rho+p_{\perp}}{\vert \hat{\Omega} \vert^{1/2}}  \label{eq:mapflu2} \\
p_r^q-p_{\perp}^q&=&\frac{p_r-p_{\perp}}{\vert \hat{\Omega} \vert^{1/2}}  \ . \label{eq:mapflu3}
\end{eqnarray}
These equations provide a unique correspondence between the two sets of scalars $\{\rho,p_r,p_{\perp}\}$ and $\{\rho^q,p_r^q,p_{\perp}^q\}$ once the RBG Lagrangian $\mathcal{L}_G$ is given. Together with the relations $u^{\mu}u_{\nu}=v^{\mu}v_{\nu}$ and $\chi^{\mu}\chi_{\nu} =\xi^{\mu}\xi_{\nu}$, this correspondence allows to write ${\Omega^\mu}_{\nu}$ in Eq.(\ref{eq:Omegafluid}) in terms of  the solution obtained in GR. Finally, the application of Eq.(\ref{eq:qggen}) allows to find the spacetime metric $g_{\mu\nu}$ which resolves the problem of RBGs coupled to an anisotropic fluid. In Sec. \ref{sec:IV} we will give an explicit example of this procedure.

\subsection{Eddington-inspired Born-Infeld gravity}

To work out an explicit and illustrative scenario for the above mapping on RBGs, let us consider here the case of the so-called Eddington-inspired Born-Infeld (EiBI) gravity, which has recently attracted a great deal of interest in the literature \cite{BIapp0,BIapp1,BIapp2,BIapp3,BIapp4,
BIapp5,BIapp6,BIapp7,BIapp8}. Its action can be conveniently expressed as
\begin{equation}\label{BIgrav}
\mathcal{S}_{EiBI}=\frac{1}{\kappa^2 \epsilon} \int d^4 x \left[\sqrt{-q}-\lambda \sqrt{-g}\right] \ ,
\end{equation}
where $q$ is the determinant of the metric $q_{\mu\nu}\equiv g_{\mu\nu}+\epsilon R_{\mu\nu}(\Gamma)$, and the parameter $\lambda$ is related to the effective  cosmological constant of the theory as $\Lambda_{eff}=\frac{\lambda-1}{\epsilon\kappa^2}$. The (length-squared) parameter $\epsilon$ controls the deviations from GR, such that for fields $\vert R_{\mu\nu} \vert \ll \epsilon^{-1}$, then GR + $\Lambda_{eff}$ is recovered. A full account of this theory, its extensions and applications can be found in the recent review \cite{BeltranJimenez:2017doy}. In this case, the deformation matrix ${\Omega^\mu}_{\nu}$ appearing in Eq.(\ref{eq:qggen}) is determined by the relation \cite{BeltranJimenez:2017doy}
\begin{equation} \label{eq:Omegadef}
\vert \hat{\Omega} \vert^{1/2} ({\Omega^\mu}_{\nu})^{-1}=\lambda {\delta^\mu}_{\nu} -\kappa^2 \epsilon {T^\mu}_{\nu} \ ,
\end{equation}
which clearly shows its dependence on the matter sources alone.

Thus, considering in this case an anisotropic fluid given by Eq.(\ref{eq:Tmunufluid0}) and an ansatz of the form (\ref{eq:Omegafluid}) for the deformation matrix ${\Omega^\mu}_{\nu}$, after a bit of algebra Eq.(\ref{eq:Omegadef}) tells us that
\begin{eqnarray}
\alpha&=&\frac{(\lambda-\tilde{p}_{\perp})(\lambda-\tilde{p}_r)^{1/2}}{(\lambda+\tilde{\rho})^{1/2}} \\
\beta&=&\frac{(\lambda-\tilde{p}_{\perp})(\lambda+\tilde{\rho})^{1/2}}{(\lambda-\tilde{p}_r)^{1/2}} \\
\gamma&=&(\lambda+\tilde{\rho})^{1/2}(\lambda-\tilde{p}_r)^{1/2} \ ,
\end{eqnarray}
where $\tilde{\rho} \equiv \epsilon \kappa^2 \rho$, $\tilde{p}_r \equiv \epsilon \kappa^2 p_r$, and $\tilde{p}_{\perp} \equiv \epsilon \kappa^2 p_{\perp}$ are the fluid functions on the RBG frame. Inserting this result into the fluid mapping equations (\ref{eq:mapflu1}), (\ref{eq:mapflu2}), (\ref{eq:mapflu3}), and after some algebra we are led to the result
\begin{eqnarray}
\lambda+\tilde{\rho}&=&\sqrt{\frac{1+ \left[\tilde{p}^q_{\perp}+\tfrac{(\tilde{\rho}^q+\tilde{p}^q_r)}{2}\right]}{1+ \left[\tilde{p}^q_{\perp} -\tfrac{(\tilde{\rho}^q+\tilde{p}^q_r)}{2} \right]}} \frac{1}{[1+\frac{(\tilde{p}^q_r-\tilde{\rho}^q)}{2}]} \label{eq:fluid1}  \\
\lambda-\tilde{p}_r&=&\sqrt{\frac{1+ \left[\tilde{p}^q_{\perp}-\tfrac{(\tilde{\rho}^q+\tilde{p}^q_r)}{2}\right]}{1+ \left[\tilde{p}^q_{\perp} + \tfrac{(\tilde{\rho}^q+\tilde{p}^q_r)}{2} \right]}}  \frac{1}{[1+\frac{(\tilde{p}^q_r-\tilde{\rho}^q)}{2}]}  \label{eq:fluid2}  \\
\lambda-\tilde{p}_{\perp}&=&  \frac{1}{\sqrt{1+\left[ \tilde{p}^q_{\perp} +\frac{(\tilde{\rho}^q+\tilde{p}^q_r)}{2} \right]}\sqrt{1+\left[ \tilde{p}^q_{\perp} -\frac{(\tilde{\rho}^q+\tilde{p}^q_r)}{2} \right]}}  \ , \label{eq:fluid3}
\end{eqnarray}
where $\tilde{\rho}^q \equiv \epsilon \kappa^2 \rho^q$, $\tilde{p}_r^q \equiv \epsilon \kappa^2 p_r^q$, and $\tilde{p}_{\perp}^q \equiv \epsilon \kappa^2 p_{\perp}^q$ are the fluid functions on the GR frame. These relations establish the explicit correspondence between the fluid functions on the GR side ($q-$superindex) and those on the EiBI side. In the next section we will provide an illustrative example of how this correspondence works by considering the case of NEDs, which naturally admit an anisotropic fluid description.

\section{NEDs as anisotropic fluids} \label{sec:III}

\subsection{General description}

Nonlinear electrodynamics theories are described by actions of the form\footnote{For simplicity we shall not consider here functions of the second field invariant $F_{\mu\nu}F^{*\mu\nu}$, where $F^{*\mu\nu}=\frac{1}{2}\epsilon^{\mu\nu\alpha\beta}F_{\alpha\beta}$ is the dual of the field strength tensor $F_{\mu\nu}$.}
\begin{equation} \label{eq:actionned}
\mathcal{S}_m=\frac{1}{8\pi}\int d^4x \sqrt{-g}\,\varphi(X) \ ,
\end{equation}
where $\varphi(X)$ is some function of the field invariant $X=-\frac{1}{2}F_{\mu\nu}F^{\mu\nu}$ (with $F_{\mu\nu}=\partial_{\mu}A_{\nu}-\partial_{\nu}A_{\mu}$ the field strength tensor, and $F^{\mu\nu}=g^{\mu\alpha}g^{\nu\beta}F_{\alpha\beta}$) characterizing the particular model. The stress-energy tensor derived from (\ref{eq:actionned}) reads
\begin{equation}\label{eq:TmnNED}
{T^\mu}_\nu =-\frac{1}{4\pi}\left[\varphi_X {F^\mu}_\alpha{F^\alpha}_\nu-\frac{\varphi(X)}{2}{\delta^\mu}_\nu \right] \ ,
\end{equation}
where $\varphi_X \equiv d\varphi/dX$. The corresponding field equations for the NED field take the form
\begin{equation}
\partial_\mu(\sqrt{-g}\,\varphi_X F^{\mu\nu})=0 \ .
\end{equation}
For static spherically symmetric  configurations this leads to a single nonzero component in the radial direction, which satisfies $\varphi_X r^2\sqrt{-g_{tt}g_{rr}}F^{tr}=Q$, with $Q$ an integration constant identified as the electric charge of the field. Given that the invariant $X$ takes the form $X=-g_{tt}g_{rr}(F^{tr})^2$, the field equations lead to $X \varphi_X^2=Q^2/r^4$, which allows to algebraically solve for $X=X(r)$ once a function $\varphi(X)$ is specified. As a result, Eq.(\ref{eq:TmnNED}) can be written as
\begin{equation} \label{eq:TmnNEDdiag}
{T}{^\mu}_{\nu}=\frac{1}{8\pi}\text{diag}\left(\varphi-2X\varphi_X,\,\varphi-2X\varphi_X,\,\varphi,\,\varphi \right) \ .
\end{equation}
A glance at Eq.(\ref{eq:Tmunufluid2}) puts forward that NEDs can indeed be seen as anisotropic fluids satisfying the relations $-\rho~=~\varphi-2X\varphi_X$, $p_r=-\rho$, and $p_{\perp}=\varphi$. Since these relations imply that $X=X(\rho)$, we can interpret $p_{\perp}$ as $p_{\perp}=K(\rho)$, where $K(\rho)$ characterizes a particular fluid model in much the same way as $\varphi(X)$ specifies a given NED.

\subsection{NEDs mapped into NEDs}

By imposing on the left-hand side of Eqs.(\ref{eq:fluid1}) and (\ref{eq:fluid2}) the NED condition (on the RBG frame) $p_r=-\rho$, and dividing the two equations, one finds that the NED condition (on the GR frame) $p_r^q=-\rho^q$ is automatically recovered. This implies that NEDs on the RBG and GR frames naturally map into each other. Note that this property was not obvious a priori given the highly nonlinear character of the mapping between theories. Denoting from now on the two sets of densities and pressures with explicit labels ``GR" and ``BI", respectively, to be as clear as possible, from Eqs.(\ref{eq:fluid1}), (\ref{eq:fluid2}) and (\ref{eq:fluid3}) one finds that the map between fluids induced by the $\text{EiBI}\to \text{GR}$ transformation takes the form (recall that tildes denote an implicit $\epsilon\kappa^2$ factor )
\begin{eqnarray}
\tilde{\rho}_{\text{\tiny BI}}&=&\frac{\lambda\tilde{\rho}_{\text{\tiny GR}}-(\lambda-1)}{1-\tilde{\rho}_{\text{\tiny GR}}} \label{mapned1} \\
\tilde{K}_{\text{\tiny BI}}&=&\frac{\lambda \tilde{K}_{\text{\tiny GR}}+(\lambda-1)}{1+\tilde{K}_{\text{\tiny GR}}} \ , \label{mapned2}
\end{eqnarray}
where $\tilde{K}_{\text{\tiny GR}}$ and $\tilde{K}_{\text{\tiny BI}}$ specify the corresponding fluid/NED models on the GR and BI sides, respectively. From the above expressions (\ref{mapned1}) and (\ref{mapned2}) it is clear that starting from Maxwell electrodynamics in any of the frames will generate a NED in the other one. The structure of the transformations above, in fact, indicates that any M\"{o}bius-type NED of the form
\begin{equation} \label{eq:KGR}
\tilde{K}_{\text{\tiny GR}}(\tilde{\rho}_{\text{\tiny GR}})=\frac{a_{\text{\tiny GR}}+b_{\text{\tiny GR}}\tilde{\rho}_{\text{\tiny GR}}}{c_{\text{\tiny GR}}+d_{\text{\tiny GR}}\tilde{\rho}_{\text{\tiny GR}}} \ ,
\end{equation}
with constant coefficients $\{a_{\text{\tiny GR}},b_{\text{\tiny GR}},c_{\text{\tiny GR}},d_{\text{\tiny GR}}\}$  will turn into another M\"{o}bius-type NED with new coefficients $\{a_{\text{\tiny BI}},b_{\text{\tiny BI}},c_{\text{\tiny BI}},d_{\text{\tiny BI}}\}$. To be precise, using Eqs.(\ref{mapned1}) and (\ref{mapned2}) this structure is transferred into the BI side as
\begin{equation} \label{eq:KBIl}
\tilde{K}_{\text{\tiny BI}}(\tilde{\rho}_{\text{\tiny BI}})=\frac{a_{\text{\tiny BI}}+b_{\text{\tiny BI}}\tilde{\rho}_{\text{\tiny BI}}}{c_{\text{\tiny BI}}+d_{\text{\tiny BI}}\tilde{\rho}_{\text{\tiny BI}}} \ ,
\end{equation}
with the following correspondences between coefficients
\begin{eqnarray}
a_{\text{\tiny BI}}&=&(\lambda-1)[(\lambda-1)d_{\text{\tiny GR}}+\lambda(b_{\text{\tiny GR}}+c_{\text{\tiny GR}})]+\lambda^2 a_{\text{\tiny GR}} \notag \\
b_{\text{\tiny BI}}&=&(\lambda-1)(c_{\text{\tiny GR}}+d_{\text{\tiny GR}})+\lambda(a_{\text{\tiny GR}}+b_{\text{\tiny GR}}) \notag \\
c_{\text{\tiny BI}}&=&(\lambda-1)(b_{\text{\tiny GR}}+d_{\text{\tiny GR}})+\lambda(a_{\text{\tiny GR}}+c_{\text{\tiny GR}})\\
d_{\text{\tiny BI}}&=&a_{\text{\tiny GR}}+b_{\text{\tiny GR}}+c_{\text{\tiny GR}}+d_{\text{\tiny GR}} \ .\notag
\end{eqnarray}
This result yields the intriguing property that the determinant of the two M\"{o}bius transformations coincide, {\it i.e.}, $\text{det}(\tilde{K}_{\text{\tiny GR}})=\text{det}(\tilde{K}_{\text{\tiny BI}})=a_{\text{\tiny GR}}d_{\text{\tiny GR}}-b_{\text{\tiny GR}}c_{\text{\tiny GR}}$. This puts forward an underlying global conformal symmetry in the mapping between the EiBI theory and GR.

The M\"{o}bius-type NED structure above is very general and encompasses most of the models known in the literature. For this family of models one finds, in particular, that the subset  that remains invariant under the mapping is not empty. Defining this set as those models for which $\{a_{\text{\tiny BI}}=a_{\text{\tiny GR}},b_{\text{\tiny BI}}=b_{\text{\tiny GR}},c_{\text{\tiny BI}}=c_{\text{\tiny GR}},d_{\text{\tiny BI}}=d_{\text{\tiny GR}}\}$, one finds two families of solutions depending on the value of $\lambda$. If $\lambda=1$, then $a_{\text{\tiny GR}}=0$ and $c_{\text{\tiny GR}}=-b_{\text{\tiny GR}}$, with $b_{\text{\tiny GR}}$ and $d_{\text{\tiny GR}}$ free parameters. If $\lambda\neq 1$ then $c_{\text{\tiny GR}}=-(a_{\text{\tiny GR}}+b_{\text{\tiny GR}})$ and $d_{\text{\tiny GR}}=a_{\text{\tiny GR}}/(\lambda-1)$. Note that Maxwell electrodynamics $K(\rho)=\rho$ is not in this subset (but $K(\rho)=-\rho$ is, though it lacks physical interest).

From now on we shall focus on asymptotically flat solutions, $\lambda=1$, for which (\ref{eq:KBIl}) becomes
\begin{equation}
\tilde{K}_{\text{\tiny BI}}(\tilde{\rho}_{\text{\tiny BI}})=\frac{a_{\text{\tiny GR}}+(a_{\text{\tiny GR}}+b_{\text{\tiny GR}})\tilde{\rho}_{\text{\tiny BI}}}{(a_{\text{\tiny GR}}+c_{\text{\tiny GR}})+(a_{\text{\tiny GR}}+b_{\text{\tiny GR}}+c_{\text{\tiny GR}}d_{\text{\tiny GR}})\tilde{\rho}_{\text{\tiny BI}}} \ .
\end{equation}
The above equations and considerations allow to reconstruct the NED model on each side of the BI/GR correspondence. Focusing our attention upon electrostatic, spherically symmetric configurations, the stress-energy tensor (\ref{eq:TmnNED}) on the GR side reads
\begin{equation} \label{eq:emNEDZ}
\bar{T}{^\mu}_{\nu}=\frac{1}{8\pi}\text{diag}(\Phi-2Z\Phi_Z,\Phi-2Z\Phi_Z,\Phi,\Phi) \ ,
\end{equation}
where $Z$ is the electromagnetic field invariant of a NED with Lagrangian density $\Phi(Z)$ coupled to $q_{\mu\nu}$, {\it i.e.},  the invariant $Z=-\frac{1}{2}B_{\mu\nu}B^{\mu\nu}$ is associated to the field strength $B_{\mu\nu}=\partial_\mu B_\nu-\partial_\nu B_\mu$ and $B^{\mu\nu}=q^{\mu\alpha}q^{\nu\beta}B_{\alpha\beta}$. Identifying the stress-energy tensor (\ref{eq:emNEDZ}) with that corresponding to the anisotropic fluid (\ref{eq:Tmunufluid2}) yields the two relations $\tilde{\rho}_{\text{\tiny GR}}=2Z\hat{\Phi}_Z-\hat{\Phi}$ and $\tilde{K}_{\text{\tiny GR}}=\hat{\Phi}$, where $\hat{\Phi} \equiv \epsilon \kappa^2 \Phi/(8\pi)$. Inserting the second relation into (\ref{eq:KGR}) yields $\tilde{\rho}_{\text{\tiny GR}}=\frac{a_{\text{\tiny GR}}-c_{\text{\tiny GR}}\hat{\Phi}}{-b_{\text{\tiny GR}}+d_{\text{\tiny GR}}\hat{\Phi}}$, and plugging this result back into the first relation one finds the result
\begin{equation} \label{eq:NEDint}
-\left(\frac{b_{\text{\tiny GR}}-d_{\text{\tiny GR}}\hat{\Phi}}{a_{\text{\tiny GR}}-(b_{\text{\tiny GR}}+c_{\text{\tiny GR}})\hat{\Phi}+d_{\text{\tiny GR}}\hat{\Phi}^2}\right)d\hat{\Phi}=\frac{dZ}{2Z} \ ,
\end{equation}
which can be readily integrated as
\begin{eqnarray} \label{eq:NEDgen}
&&\frac{(-b_{\text{\tiny GR}}+c_{\text{\tiny GR}})\arctan[\frac{-b_{\text{\tiny GR}}-c_{\text{\tiny GR}}+2d_{\text{\tiny GR}}\hat{\Phi}}{D^{1/2}}]}{D^{1/2}}\\
&+&\frac{1}{2}\log[a_{\text{\tiny GR}}-(b_{\text{\tiny GR}}+c_{\text{\tiny GR}})\hat{\Phi}+d_{\text{\tiny GR}}\hat{\Phi}^2]=\frac{1}{2}\log\left[\frac{Z}{Z_0}\right] \ , \nonumber
\end{eqnarray}
(provided that $D \equiv 4a_{\text{\tiny GR}}d_{\text{\tiny GR}}-(b_{\text{\tiny GR}}+c_{\text{\tiny GR}})^2>0$), where $Z_0$ is an integration constant. The above expression allows for a resolution of the function $\hat{\Phi}(Z)$ of the GR side, once the corresponding function on the EiBI side, as given by the coefficients $\{a_{\text{\tiny BI}},b_{\text{\tiny BI}},c_{\text{\tiny BI}},d_{\text{\tiny BI}}\}$, is specified.
As for the invariant subset of NEDs with $\lambda=1$, $a_{\text{\tiny BI}}=0$, and $c_{\text{\tiny GR}}=-b_{\text{\tiny GR}}$, the above result is singular.
Direct integration of Eq.(\ref{eq:NEDint}) in that case yields the analytical expression
\begin{equation}
\hat{\Phi}(Z)=-\frac{b_{\text{\tiny GR}}}{d_{\text{\tiny GR}}\text{ProductLog}\left[-\frac{b_{\text{\tiny GR}}}{d_{\text{\tiny GR}}\sqrt{\frac{Z}{Z_0}}}\right]}
\end{equation}
which is well defined provided that $b_{\text{\tiny GR}}/d_{\text{\tiny GR}}<0$.

The mapping presented in this work is particularly transparent for those models where the GR solution is known in closed, analytical form. This is precisely the case for spherically symmetric, electrovacuum solutions out of NEDs. Indeed, in this case, for asymptotically flat solutions with a Maxwell fall-off at infinity, $X_{\text{Maxwell}}=Q^2/r^4$, the general solution is given by \cite{DiazAlonso:2009ak}
\begin{eqnarray}
ds_{GR}^2&=&-C(x)dt^2+\frac{dx^2}{C(x)}+x^2 d\Omega^2 \label{eq:linelesss} \\
C(x)&=&1-\frac{2M(x)}{x} \label{eq:metricGR}  \\
M(x)&=&M_0+\frac{\kappa^2}{2} \int_x^{\infty} x^2 {T^t}_t(x) dx  \label{eq:massGR}
\end{eqnarray}
where $d\Omega^2=d\theta^2 + \sin^2(\theta)d\phi^2$ is the angular element in the two-spheres, $M_0$ is Schwarzschild mass, and ${T^t}_t=-\rho_{\text{\tiny GR}}$. This expression leads to an immediate computation of the metric function when a given model is specified, allowing to generate the plethora of solutions already known for many NED models satisfying the conditions above.

\section{An example: Maxwell electrodynamics} \label{sec:IV}

To illustrate the method explained in the previous section, here we shall derive the solution for the case of Maxwell electrodynamics coupled to EiBI gravity. Maxwell Lagrangian corresponds to $\varphi(X)=X$, which leads simply to $K_{\text{\tiny BI}}=\rho_{\text{\tiny BI}}$. Assuming asymptotically flat solutions ($\lambda=1$), the fluid mapping equations (\ref{mapned1}) and (\ref{mapned2}) yield the corresponding functions on the GR side as
\begin{eqnarray}
\tilde{\rho}_{\text{\tiny GR}}&=&\frac{\tilde{\rho}_{\text{\tiny BI}}}{1+\tilde{\rho}_{\text{\tiny BI}}} \\
\tilde{K}_{\text{\tiny GR}}&=&\frac{\tilde{\rho}_{\text{\tiny GR}}}{1-2\tilde{\rho}_{\text{\tiny GR}}} \ , \label{eq:KMax}
\end{eqnarray}
To avoid overcharging the notation, in what comes next we will drop the label ``GR" and thus all functions will be implicitly assumed to be computed on the GR side (unless explicitly stated).
From the matter conservation equation, $\nabla_{\mu}^{(g)}T^{\mu\nu}=0$, in a static spherically symmetric background (\ref{eq:linelesss}), one finds that for a NED-type fluid this equation can be expressed as
\begin{equation}
\frac{d\rho}{dx}+\frac{2[\rho+K(\rho)]}{x}=0 \ ,
\end{equation}
and can be suitably rearranged as
\begin{equation}
x^2=x_0^2 \exp \left[-\int^{\rho} \frac{d\rho'}{\rho'+K(\rho')} \right] \ .
\end{equation}
where $x_0$ is an integration constant. Inserting the expression (\ref{eq:KMax}) into the above equation and performing the integral one arrives to the result
\begin{equation}
\rho(1-\epsilon \kappa^2 \rho)=\frac{x_0^4}{x^4} \ .
\end{equation}
Imposing on this relation the asymptotic Maxwell limit,  $\rho(x) \underset{x \rightarrow \infty}{\longrightarrow} {Q}^2/(8\pi x^4)$, allows to fix the constant $x_0^4={Q}^2/(8\pi)$, so that the quadratic equation above is solved as (choosing the branch with asymptotic Maxwell limit)
\begin{equation} \label{eq:density}
\rho=\frac{1-\sqrt{1-\frac{\epsilon \kappa^2 {Q}^2 }{2\pi x^4}}}{2  \epsilon \kappa^2} \ .
\end{equation}
From Eq.(\ref{eq:massGR}) this implies that
\begin{equation} \label{eq:mf1}
M_{x}=\frac{\kappa^2 x^2}{2} \rho=\frac{x^2}{4\epsilon} \left(1-\sqrt{1-\frac{\epsilon \kappa^2 {Q}^2 }{2\pi  x^4}} \right) \ .
\end{equation}
where $M_x \equiv dM/dx$.

It is instructive to find the explicit form of the NED Lagrangian density generating this expression. Indeed, from the general solution (\ref{eq:NEDgen}) particularized to this case, and by demanding the recovery of Maxwell electrodynamics for small fields, one gets
\begin{equation} \label{eq:NEDinverse}
\hat{\Phi}(Z)=\frac{1}{2}\left(-1+\sqrt{1+\frac{4Z}{Z_0}}\right) \ ,
\end{equation}
which also fixes the value of the integration constant $Z_0$ as $Z_0^{-1}=\epsilon \kappa^2/(8\pi)$. The square-root structure of this Lagrangian density is actually the same as that of Born-Infeld electrodynamics \cite{BI34}, defined by the Lagrangian density\footnote{From the definitions introduced so far this identification is exact provided that $\beta^2 =-2\pi/(\epsilon\kappa^2 )$, \emph{i.e.}, for the negative branch of $\epsilon$. For the positive branch of $\epsilon>0$ this Lagrangian flips the sign inside the square-root and a global sign out of it as compared to Born-Infeld electrodynamics.}
\begin{equation}\label{eq:BIL}
\mathcal{L}_{BI}=2\beta^2\left(1-\sqrt{1-\frac{X}{\beta^2}}\right) \ ,
\end{equation}
(where $\beta$ is Born-Infeld parameter) originally introduced to bound both the electric field and the self-energy of a point-like charge. Indeed, solving the NED equations in this case, $\nabla_{\mu}(\Phi_Z B^{\mu\nu})=0$, which for static, spherically symmetric solutions read $x^2 \Phi_Z Z^{1/2}={Q}$, we obtain the field invariant
\begin{equation} \label{eq:Einverse}
Z(x)=\frac{{Q}^2}{x^4}\frac{1}{\left(1-\frac{\epsilon\kappa^2}{2\pi}\frac{{Q}^2}{x^4}\right)} \ ,
\end{equation}
For the correct Born-Infeld electrodynamics branch, $\epsilon<0$, this expression yields a bounded electric field at the center, $Z(x=0)= \beta^2= \vert 2\pi/(\epsilon \kappa^2) \vert$, which is to be expected from the known behaviour of this model. In addition, starting from the NED stress-energy tensor (\ref{eq:emNEDZ}) and after a bit of algebra, one arrives at the same equation for the energy density  (\ref{eq:density}) as obtained by direct application of our method, which confirms the consistence of the approach. The integration of that equation throughout all space reveals that the boundness of the field invariant is transferred into the boundness of the total energy of the electromagnetic field, as it should be expected\footnote{In the branch $\epsilon>0$ the field invariant diverges instead at a radius $x_{\star}^4=\epsilon \kappa^2 {Q}^2/(2\pi)$, but both the energy density and the total energy turn out to be  finite as well.}.

\smallskip The next step of the mapping is to generate the solution for the metric functions on the EiBI gravity side starting from its GR counterpart above. In order to make contact with relevant previous literature, we will focus on the negative, Born-Infeld branch. For convenience, we introduce a length scale as $\epsilon=-2l_{\epsilon}^2<0$, define the charge radius $r_{Q}^2 \equiv \kappa^2 {Q}^2/(4\pi)$, and introduce another length scale $r_c^4 \equiv l_{\epsilon}^2r_{Q}^2$, which allows to use the dimensionless variable $y=x/r_c$, in terms of which the mass function (\ref{eq:mf1}) reads
\begin{equation}  \label{eq:mf2}
M_y=\frac{r_c^3}{8l_{\epsilon}^2} y^2 \left(\sqrt{1+\frac{4}{y^4}}-1\right) \ .
\end{equation}
By integrating this expression one has fully specified the GR line element according to the set of equations (\ref{eq:linelesss}), (\ref{eq:metricGR}) and (\ref{eq:massGR}). However, we do not need to follow that path. Instead, we note that the mapping from this GR solution to the EiBI theory involves the deformation matrix ${\Omega^\mu}_{\nu}$ introduced in Eq.(\ref{eq:Omegadef}). Due to this equation, the structure in $2 \times 2$ blocks of the stress-energy tensor ${T^\mu}_\nu$ is transferred into a similar structure for this matrix as (using the anisotropic fluid representation)
\begin{eqnarray}
{\Omega^\mu}_{\nu}&=&
\left(
\begin{array}{cc}
\Omega_{1}\hat{I} &   \hat{0} \\
\hat{0} & \Omega_{2 }\hat{I}   \\
\end{array}
\right) \ ,  \label{eq:em}
\end{eqnarray}
with $\Omega_1=\lambda-\tilde{\rho}_{\text{\tiny BI}}$ and $\Omega_2=\lambda+\tilde{\rho}_{\text{\tiny BI}}$, where $\hat{I}$ and $\hat{0}$ are the $2 \times 2$ identity and zero matrices, respectively.  Using (\ref{mapned1}) and (\ref{eq:density}), the dependence of ${\Omega^\mu}_{\nu}$ on the dimensionless radial variable $y$ is completely specified, leading to
\begin{equation}
\tilde{\rho}_{\text{\tiny BI}}=\frac{y^2-\sqrt{y^2+4}}{y^2+\sqrt{y^2+4}} \ .
\end{equation}
Writing now the line element associated to $g_{\mu\nu}$ as\footnote{Though in spherically symmetric systems the gauge can always be fixed to have only two independent functions, nonetheless we shall use this form of the line element for the sake of this computation.}
\begin{equation} \label{eq:lineBI}
ds_{\text{\tiny BI}}^2=-A(r)dt^2+B^{-1}(r)dx^2+r^2(x)d\Omega^2 \ ,
\end{equation}
and using the fundamental relation between metrics, Eq.(\ref{eq:qggen}), an explicit expression for $g_{\mu\alpha}$ is automatic. Focusing first on the spherical sector and using the dimensionless variable $z=r/r_c$, we find
\begin{equation} \label{eq:zofydif}
y^2=z^2 \Omega_2=z^2(1+\tilde{\rho}_{\text{\tiny BI}}) \ ,
\end{equation}
which leads explicitly to
\begin{equation}\label{eq:zofy}
z^2=\frac{y^2+\sqrt{4+y^2}}{2} \ .
\end{equation}
Now, taking a derivative on the relation above yields $dy/dz=\Omega_1/\Omega_2^{1/2}$ and, replacing in the GR-expression (\ref{eq:mf1}), yields the result
\begin{equation} \label{eq:Mz}
M_z=\delta_1\frac{\Omega_1}{4z^2\Omega_2^{1/2}}=\delta_1\frac{z^4+1}{4z^4\sqrt{z^4-1}} \ ,
\end{equation}
where we have isolated all the constants of the problem into the parameter $\delta_1\equiv \sqrt{r_{Q}^3/l_{\epsilon}}$.
To complete the correspondence one just needs, in addition to the relation between radial coordinates (\ref{eq:zofydif}), to work out the relation (\ref{eq:qggen}) in the temporal and radial sectors, to cast the line element (\ref{eq:lineBI}) into the convenient form
\begin{equation} \label{eq:lecan}
ds^2=-\frac{C(y)}{\Omega_1}dt^2+\frac{dy^2}{\Omega_1 C(y)}+z^2(y)d\Omega^2
\end{equation}
where from (\ref{eq:metricGR}) and (\ref{eq:zofydif}) one has $C(y)=1-2M(z)/(z\Omega_1^{1/2})$, with the mass function defined in (\ref{eq:Mz}). This finally closes the problem.

The line element (\ref{eq:lecan}) with the expressions (\ref{eq:zofy}) and (\ref{eq:Mz})  exactly match those found in previous works \cite{Olmo:2013gqa,Olmo:2012nx} (see also the discussion of \cite{Olmo:2015bya} for the proper interpretation of this line element) by direct integration of the differential equations of EiBI gravity coupled to a Maxwell field. The derivation here only involved the resolution of the GR equations coupled to a specific NED (which turned out to correspond to Born-Infeld electrodynamics) and then some purely algebraic manipulations. The analysis of the structure of this line element reveals the existence of geometries replacing the point-like central singularity by a wormhole, which provides geodesically complete solutions with a non-singular character regarding the paths of physical observers and the scattering of waves \cite{Olmo:2015dba}. The presence of the wormhole structure is inferred from the analysis of Eq.(\ref{eq:zofy}), which implies that $z(y)$ has a minimum at $y=0$, where it bounces off. The bottom line of this result is that, starting from a known GR solution corresponding to some nonlinear matter field, the mapping allows to find the corresponding solution on the RBG side (EiBI in this example) coupled to another nonlinear field. Moreover, it illustrates how this is achieved by solving purely algebraic equations rather than differential ones.

\subsection{Solving an old puzzle}

The above correspondence allows to explain one striking result that went unexplained in the original publications \cite{Olmo:2011np,Olmo:2012nx}, were the above non-singular solutions were first found. In such publications it was considered the case of quadratic gravity defined by a Lagrangian density $\mathcal{L}_G=R+l_{P}(R^2+aR_{\mu\nu}R^{\mu\nu})$, where $l_P \equiv \sqrt{\hbar G/c^3}$ is Planck's length and $a$ a dimensionless constant, coupled to an electromagnetic Maxwell field. It was noted there that, when setting $\delta_1=\delta_c$, where the constant $\delta_c \approx 0.57206$ arises in the integration of the mass function (\ref{eq:Mz}), then a family of solutions with the mass spectrum
\begin{equation}\label{eq:mass1}
M=n_{BI}\left(\frac{N_q}{N_c}\right)^{3/2}m_P \ ,
\end{equation}
is found, where $N_q=q/e$ is the number of charges (and $e$ is proton's charge), $m_P=\sqrt{\hbar G/c}$ is Planck's mass, and $N_q^c=\sqrt{2/\alpha_{em}}$ (with $\alpha_{em}$ the fine-structure constant), defines a critical charge. The puzzle about this result lies on the fact that, despite starting with Maxwell electrodynamics, the constant $n_{BI}=\pi^{3/2}/(3\Gamma[3/4])^2 \approx 1.23605$ appearing in (\ref{eq:mass1}) is a number which arises in the computation of the total energy of the electrostatic solutions of Born-Infeld electrodynamics, derived from (\ref{eq:BIL}) as $\varepsilon_{BI}=4\pi n_{BI} q^{3/2}\beta^{1/2}$. Now, the mass spectrum (\ref{eq:mass1}) defines a set of objects with peculiar properties as compared to those with masses above or below this value (see  \cite{Olmo:2011np,Olmo:2012nx} for details), in that in this case curvature divergences go away everywhere\footnote{Nonetheless, subsequent research has shown that, regardless of whether curvature divergences are present or not, any electrovacuum solution in EiBI/quadratic gravity is non-singular, see \cite{Olmo:2015dba,Olmo:2016fuc} for details.}. The underlying reason of why a specific number so tightly related to Born-Infeld electrodynamics appears in the context of quadratic gravity coupled to Maxwell electrodynamics triggered additional research.

A first element to resolve this puzzle was provided in Ref.\cite{Olmo:2013gqa}. There it was found that, for fields whose stress-energy tensor have the same algebraic structure as that of an electrostatic, spherically symmetric field, the corresponding solutions of quadratic gravity and those of EiBI gravity (\ref{BIgrav}) are exactly the same. Indeed, if we set again $\epsilon=-2l_{\epsilon}$ then the mass (\ref{eq:mass1}) in EiBI theory reads
\begin{equation}
M=n_{BI}\left(\frac{N_q}{N_c}\right)^{3/2}m_P\left(\frac{l_P}{l_{\epsilon}}\right)^{1/2} \ ,
\end{equation}
so a new term on the ratio $l_P/l_{\epsilon}$ is picked up. We see again the re-appearance of Born-Infeld number, $n_{BI}$, in this case. Now, the explicit example of the mapping above provides the second element to resolve this puzzle. Indeed, if we compute the total energy associated to the Born-Infeld NED field defined by (\ref{eq:NEDinverse}), using (\ref{eq:density}) we find
\begin{equation}
\varepsilon=4\pi \int_0^{\infty} dx x^2 \rho(x)=n_{BI}q^{3/2}(8l_{\epsilon}^2)^{-1/4} \ ,
\end{equation}
and we see that this is exactly the total energy of the matter Born-Infeld field, modulo some suitable identification between the respective length scales on the matter, $\beta^2$, and gravity, $l_{\epsilon}^2$, sides (see footnote 4 above).  The bottom line of this discussion is that, via the correspondence described in the present work, the nonlinear properties of the matter fields coupled to GR are somewhat transferred to the gravitational sector on the RBG side of the mapping which, for the case of the puzzle described here, involves two stages: GR + Born-Infeld NED $\rightarrow$ quadratic gravity + Maxwell $\rightarrow$ EiBI gravity + Maxwell, making the Born-Infeld NED number $n_{BI}$ to emerge on the RBG side, as well as the square-root structure of the respective Lagrangian densities.

\section{Conclusion and perspectives}\label{sec:V}

In this work we have introduced a correspondence between the space of solutions of Ricci-based theories of gravity formulated in the metric-affine approach, and that of General Relativity. This is possible thanks to the formulation of the RBG field equations in the Einstein frame in terms of  an auxiliary metric, with the matter fields sourcing the right-hand side of such equations via nonlinear contributions. The existence of this mapping is independent of any assumption on symmetries of the scenario and/or the solutions under consideration, being instead completely general.

We have illustrated this mapping by explicitly formulating it for anisotropic fluids, showing the correspondence between the spaces of solutions of Eddington-inspired Born-Infeld gravity and GR coupled to different shapes of this same matter source. Moreover, we have used this correspondence together with the fact that spherically symmetric, non-linear, electric fields can be seen as a particular kind of anisotropic fluid, to construct the EiBI electrovacuum solutions in terms of the corresponding solutions in the GR frame. We also showed that there exists a family of NEDs, with the form of a M\"{o}bius transformation which, under the map that relates GR and EiBI, maintains the structure of the NED. This suggests that these mappings between theories may hide new symmetries that are to be explored.

As an explicit application of the map, we found that when EiBI is coupled to Maxwell electrodynamics, the corresponding matter theory on the GR side is a specific NED with a square-root structure, which for the negative branch of the EiBI parameter exactly coincides with that of Born-Infeld theory of electrodynamics. This allowed us to fully reconstruct known solutions in the literature of EiBI gravity, using the mapping instead of solving differential field equations and, as a bonus, to solve an old puzzle related to the appearance of a specific number associated to the Born-Infeld NED, in the gravitational sector of EiBI gravity coupled to Maxwell electrodynamics. This explicit example illustrates how the mapping introduced in this work may breath new life into nonlinear matter models in the framework of GR regardless of their intrinsic physical interest, since their counterpart on the RBG side may admit a natural motivation. Though this work has focused on EiBI gravity, it is applicable to any other RBG, for instance, $f(R)$ theories \cite{Olmo:2011ja,Bambi:2015zch}, following a similar procedure as the one described here.

The method presented here opens a new door to attack realistic astrophysical and cosmological scenarios in RBGs, which were previously unaccessible due to the difficulty to explicitly resolving the field equations (\ref{eq:GmnGeneral}) and inverting the relation (\ref{eq:qggen}) between metrics. From this starting point, regular solutions,  gravitational waves, signatures of horizonless compact objects, less symmetric cosmological settings, and so on, in RBG theories, can now be tackled from a different perspective using the full capability of the analytical and numerical methods developed within the GR framework.

In this sense, there are several specific cases of interest for which this method can prove its power. We underline here the case of the axisymmetric solutions of the RBG field equations. This is a problem of enormous interest from both a theoretical and phenomenological perspective and, as such, different attempts to find explicit solutions in modified theories of gravity have been proposed in the literature (like the Janis-Newman algorithm \cite{NJmethod,CLS,NEDstructure9,Erbin}). Having obtained in this work the electrovacuum, static, spherically solution of the EiBI gravity by using the mapping, the next step would be to find the counterpart of the Kerr-Newman solution in several RBGs following a similar approach \cite{MOR18}. This would allow us to directly explore normal modes, perturbations of all kinds, black hole shadows, the existence of echoes in gravitational wave emission,\ldots, and look for observational discriminators with respect to GR predictions, something hardly accessible by other means within the context of these theories.

From a  more technical point of view, the study of more general setups involving time-dependent electromagnetic fields without a correspondence with fluids is an open problem that we hope to address in the future. Similarly, the extension of the analysis presented here to other matter sources of interest for astrophysics and cosmology, such as scalar or non-abelian fields, will also be discussed in detail elsewhere. Work along several of the lines above is currently underway.

\section*{Acknowledgements}

GJO is funded by the Ramon y Cajal contract RYC-2013-13019 (Spain). DRG is funded by the Funda\c{c}\~ao para a Ci\^encia e a Tecnologia (FCT, Portugal) postdoctoral fellowship No.~SFRH/BPD/102958/2014 and by the FCT research grants No. UID/FIS/04434/2013 and No. PTDC/FIS-OUT/29048/2017. This work is supported by the Spanish projects FIS2014-57387-C3-1-P, FIS2017-84440-C2-1-P (AEI/FEDER, EU), the project H2020-MSCA-RISE-2017 Grant FunFiCO-777740, the project SEJI/2017/042 (Generalitat Valenciana), the Consolider Program CPANPHY-1205388, and the Severo Ochoa grant SEV-2014-0398 (Spain).  This article is based upon work from COST Action CA15117, supported by COST (European Cooperation in Science and Technology).

\end{document}